\documentclass[english,aps,preprint]{revtex4}
\usepackage[T1]{fontenc}
\usepackage[latin1]{inputenc}

\makeatletter

\usepackage{babel}
\makeatother
\begin{document}

\title{Experimental Characterization of 1-D Velocity Selection}

\author{S.H. Myrskog, J.K. Fox, A.M. Jofre$^{1}$, L.R. Segal, S.R. Mishra$^{2}$
and A.M. Steinberg}

\affiliation{Dept. of Physics, University of Toronto, Canada \\
 $^{1}$ National Institute of Standards and Technology, Gaithersburg,
USA \\
 $^{2}$ Center for Advanced Technology,\ Indore 152013, India}

\date{January 21, 2005 }


\begin{abstract}
We demonstrate a 1-D velocity selection technique which relies on
combining magnetic and optical potentials. \ We have selected atom
clouds with temperatures as low as 2.9\% of the initial temperature,
with an efficiency of 1\%. \ The efficiency (percentage of atoms
selected) of the technique can vary as slowly as the square root of
the final temperature. \ In addition to selecting the coldest atoms
from a cloud, this technique imparts a sharp cut-off in the velocity
distribution. \ The cold selected atoms are confined in a small well,
spatially separated from higher energy atoms. Such a non-thermal distribution
may be useful for atom optics experiments, such as studies of atom
tunneling. 
\end{abstract}
\maketitle

\section{Introduction}

The field of atom optics has experienced a period of dramatic growth
recently, largely due to the success of experiments on Bose-Einstein
Condensation (BEC). \ The achievement of BEC itself was based largely
on the success of evaporative cooling\cite{evap,evap2}. Unfortunately,
evaporative cooling is usually accomplished with a very high cost
in atom number, such that the typical efficiency (percentage of atoms
remaining of the initial sample) varies linearly with temperature\cite{BEC-amo}.
\ While this method works very well, it is quite lossy compared to
certain other techniques, particularly if one is concerned with temperatures
along only one dimension. \ Other cooling techniques have been shown
to be useful for achieving high phase-space density, such as Raman
Sideband Cooling\cite{RamanSide1,RamanSideband}. \ This technique
relies on deep potential wells, which imply a large zero-point energy,
typically resulting in temperatures on the order of the recoil temperature
after adiabatic release. \ To cool below the recoil limit (in 1-D)
methods such as Raman cooling \cite{RamanCool1,RamanCool} and VSCPT\cite{VSCPT}
have proven useful in getting cold atoms, but again they prove inefficient
in atom number.

In this paper we describe a technique to select the coldest atoms
in one dimension to obtain long deBroglie wavelengths which could
be used to study atom-optical effects such as tunneling\cite{Superlattice},
quantum potential scattering\cite{Muga} and chaos\cite{Raizen,ChaosTunnel,Ammann}.
\ This technique can be implemented after the earlier listed techniques
as it has no fundamental limit. \ The process of velocity selection
can also be quite efficient, the efficiency of the selection dropping
only as the square root of the ratio of the final to initial temperature.
\ This selection process confines the atoms in a single potential
well and imparts a sharp cut-off in velocity, yielding a non-thermal
distribution. \ The sharp cut-off in energy guarantees that there
are no high energy atoms in the system, making it useful for studying
effects sensitive to energy, such as tunneling and collisions.

\section{Theory}

We begin by considering a dilute cloud of laser-cooled atoms where
the density is low enough that collisions and rethermalization may
be neglected. \ Selection of low energy atoms is accomplished by
sudden turn on of a potential that has a local minimum. \ The potential
is comprised of a potential gradient and a barrier arranged to create
a local minimum in the region of the atom cloud as shown in Figure
1. \ In the classical limit any atom that begins in the potential
well and has sufficiently low kinetic energy will become trapped in
the well. \ Higher energy atoms will simply pass over the barrier
and accelerate along the potential gradient, quickly becoming spatially
separated from the trapped atoms. \ 

We first consider the limit where the potential energy difference
across the cloud due to the gradient ($\approx$ $2U^{\prime}r$ ,where
$r$ is the rms radius of the cloud) is negligible compared to the
kinetic energy ($k_{B}T/2$ at an initial temperature $T$). \ Since
the initial distribution of laser-cooled atoms is essentially thermal,
the final velocity distribution is a truncated Gaussian\textbf{,}
keeping the low-velocity atoms and discarding the high-energy atoms.
\ The efficiency of the selection process is then $\eta=\frac{1}{\sqrt{2\pi}\sigma}\int_{-v_{m}}^{v_{m}}e^{-v^{2}/2\sigma^{2}}dv$
where $\sigma$ is the rms velocity of the cloud and $v_{m}$ is the
maximum velocity a selected atom can have. \ The Gaussian distribution
is relatively flat for $v_{m}\ll\sigma$, allowing the efficiency
to be approximated as $\eta\simeq\sqrt{\frac{2}{\pi}}\frac{v_{m}}{\sigma}=\sqrt{\frac{4U_{0}}{\pi k_{B}T}}.$ \ The
selected cloud now has an essentially square velocity distribution,
with a mean kinetic energy of $U_{0}/3.$ If we define this mean kinetic
energy as a pseudo-temperature $k_{B}T_{s}/2$, then the efficiency
of selection is $\eta=\sqrt{\frac{6T_{s}}{\pi T}}$. This allows one
to reduce the mean (1D) kinetic energy by a factor of 200 while still
retaining 10\% of the atoms. \ \ In comparison, the efficiency of
evaporative cooling depends linearly on the temperature ratio\cite{BEC-amo},
implying that about 0.5\% of the atoms would remain after cooling
by a factor of 200.

In the case of a high gradient, the atoms begin with a significant
amount of potential energy as compared to the height of the barrier.
\ \ Since atoms farther away from the barrier gain more kinetic
energy than closer atoms, there is a limit to the distance an atom
can be from the barrier and still be selected. \ The maximum distance
an atom may be from the well and still be trapped is then $z_{m}=U_{0}/U^{\prime}$.
\ \ A large gradient will make the distance $x_{m}$ smaller than
the size of the cloud. \ The most efficient selection will then occur
when the potential barrier is placed in the center of the atom cloud.\ For
$z_{m}$ and $v_{m}$ smaller than the rms spatial and velocity distributions,
the mean kinietic energy $k_{B}T_{s/2\text{ }}$works out to be $U_{0}/4$.
\ The efficiency varies as $\left(\frac{T_{s}}{T}\right)^{3/2}$ \ where
a factor $T_{s}^{1/2}$ comes from the velocity distribution and a
factor of $T_{s}$ from the gradient\cite{Fox}. The $3/2$ power
leads to a large decrease in the selection efficiency, making large
potential gradients undesirable.

Due to experimental constraints, the potential gradient cannot be
made arbitrarily small. \ In order to obtain the spatial separation
of the hot and selected clouds within a reasonable time, a minimum
gradient is required to ensure that the atom cloud propagates further
than it expands. \ When very cold atoms are desired one must always
use very small barriers, in which case the potential energy across
the cloud will be greater than the height of the barrier. For these
small barriers heights, one will have a resulting efficiency that
varies as $T_{s}^{3/2}$. \ 

In both limits of the selection process, the atoms have a maximum
energy defined by the barrier. \ No atom with energy greater than
$U_{0}$ can remain within the well created by the potential barrier
and gradient. \ The sharp cut-off in velocity (energy) can be observed
in the long-time free expansion of the cloud as a lack of the long
tails, or it can be observed directly by performing a deconvolution
of the final spatial distribution. \
For the limit of very low potential energy the velocity distribution
will approach a truncated Gaussian. \ For large gradients the velocity
distribution appears more triangular than square. \ Detailed simulations
of this technique are to be published seperately\cite{Fox}. \ Quantum
effects (preparation of single bound states) have also been predicted,
but for the regime of the present experiment, the classical approximation
is accurate.

\section{Experiment}

We begin with a cloud of $^{85}Rb$ atoms cooled in a $\sigma^{+}\sigma^{-}$
molasses to a temperature on the order of $10\mu K$. The cloud has
an rms radius $r$ of about 500 microns for this experiment. Immediately
following a cooling stage, we optically pump the atoms to the doubly
polarized $\left|F=3,m_{F}=3\right\rangle $ state to ensure that
all the atoms experience the same potential. A weak magnetic quadrupole
trap consisting of a pair of anti-Helmholtz coils is turned on in
conjunction with a pair of bias coils along the same axis as the quadrupole
coils to create potential gradients of $10$ $G/cm$ or less. The
bias coils are in a Helmholtz configuration and create a constant
magnetic field in the region of the atoms. This bias field serves
to displace the center of the quadrupole field by a distance $z_{0}=B_{b}/B^{^{\prime}}$
where $B_{b}$ is the strength of the bias field, and $B^{^{\prime}}$
is the gradient of the quadrupole field along the $z$ direction.
The magnitude of the bias field is always chosen such that the trap
center is displaced by a distance greater than the size of the cloud.
\
An optical dipole potential is turned on at the same instant as the
magnetic fields. \ The dipole barrier consists of a Ti:Sapphire laser
operating 210 GHz to the blue of the Rubidium D2 resonance. \ The
beam is focused to a vertical line 12 microns wide and 6 mm tall that
intersects the cloud of atoms. \ The intensity of the beam is controlled
with an acousto-optic modulator, enabling us to vary the potential
height of the barrier from $100\mu K$ down to the nanoKelvin range.
\ The ratio between the potential and kinetic energy is controlled
by varying the temperature and size of the initial cloud, and varying
the strength of the gradient. \ To study a larger range of parameters
the trapped atoms are sometimes compressed prior to the experiment,
heating the cloud to about $26\mu K$ while reducing its size to 160
$\mu m$. \ \ \ 

When the displaced quadrupole field is turned on, the atoms are located
far from the $z=0$ origin of the magnetic trap and instead of experiencing
a conical potential (which would couple different spatial degrees
of freedom) they see the potential locally as a (parabolic) conic
section along $x$ and $y$. \ The atoms therefore experience a potential
that is essentially separable: harmonic along the $x$ and $y$ axes,
and linear along $z.$ \ This helps to prevent mixing between the
different degrees of freedom, justifying the treatment of this as
a 1-D problem and the possibility of achieving $T_{z}\ll T_{x},T_{y}$.
\ The combination of the magnetic potential and the dipole potential
creates a well in which atoms may be trapped. \ The atoms begin accelerating
along the $z$ axis toward the center of the trap. \ Any atom which
starts off on the far side of the barrier relative to the trap center
can be trapped if the atom has sufficiently low energy.\ \ The well
depth $U_{0}$ is smaller than the barrier height by a correction
term on the order of $U^{\prime}\sigma_{L}$ where $U^{\prime}$ is
the gradient of the magnetic field and $\sigma_{L}$ is the $1/e^{2}$
radius of the laser beam. \ For large barrier heights this effect
is negligible.\ 

The atom cloud is allowed to evolve within the potential long enough
for several oscillations to occur. \ For typical parameters used
the well is 120 $\mu m$ wide, with the selected atoms having an rms
velocity of 1 $cm/s$ allowing for almost 3 full oscillations within
20 milliseconds. \ After this time, both the magnetic and optical
fields are turned off to determine the temperature by a time of flight
measurement. \ The mean energy of the cloud is determined by fitting
the clouds to a 1-D Boltzmann distribution.\ The efficiency of the
selection (fraction of atoms selected) is determined by measuring
the ratio of selected to non-selected atoms 0.5 milliseconds after
the gradient and barrier are turned off. \ A sample image of the
total cloud distribution (selected and non-selected) is shown in Figure
2, after 20 ms of free expansion, long enough for significant separation
of the selected and non-selected components.\ The different regimes
for the initial cloud are achieved by varying the trapping procedure
and MOT parameters. \ The large potential energy regime is achieved
by using a large (400 micron) cloud of atoms cooled to about 26 $\mu K$
placed into a relatively high gradient of 8 G/cm, resulting in the
potential energy across the cloud varying by 42 $\mu K$. \ The low
potential energy regime is achieved by compressing the MOT before
beginning selection. \ In this case the initial clouds are 160 microns
across with\ a temperatures near 25 $\mu K$ and are subsequently
placed into traps with gradients of 3 G/cm, with the potential energy
varying across the cloud by 6 $\mu K$ . \ Figure 3 shows the temperature
of the selected cloud versus the potential height of the barrier.
\ The initial conditions of the clouds are characterized by\ a parameter
$\beta=2U^{\prime}r/k_{B}T$, which is the ratio of the potential
energy difference across the cloud to the kinetic energy at the start
of the selection . \ \ The temperature of the selected clouds is
plotted versus the height of the potential barrier for two sets of
data with $\beta=0.23$ and $1.7$ for the solid and hollow circles
respectively. \ For the four lowest barrier height data points for
each set, the slopes are 0.34 and 0.40 respectively, in rough agreement
with the prediction of $k_{B}T_{s}=U/3$. \
The lowest cooling ratio $T_{s}/T$ we observed was $1/35$, with
the lowest temperature observed being 750 nK. \ The observed temperature
is higher than the theoretical prediction of a 1-D classical simulation
that includes only the initial cloud parameters, the barrier height
and the gradient\cite{Fox}. \ The experimental temperatures may
be higher than theory due to residual mixing of the spatial degrees
of freedom, heating by scattering photons from the dipole barrier
or leakage of atoms around the barrier.

The efficiency of the selection process is determined by measuring
the ratio of selected atoms to the total number of atoms $0.5$ milliseconds
after the selection process is completed. \ In Figure 4, we see the
efficiency for different barrier heights with the corresponding theoretical
curves. \ The initial cloud in each case started at a temperature
of $26\,\,\mu K$ but with two values of $\beta$ \ as before: $0.23$
and $1.72$, corresponding to the circles and squares respectively.
\ For $\beta=0.23$ (circles), the near square root dependence of
efficiency on well depth ($U_{o})$ is visible. \ For the higher
ratio $\beta=1.7$, a more linear dependence is observed. \ The experimental
efficiency is found to be somewhat lower than the theoretical expectation.
\ Even in the case of weak gradients, a 3/2 power law dependence
will be observed for extremely low barrier heights, since the barrier
potential will then be lower than the potential energy across the
cloud.

\section{Conclusion}

In conclusion, we have demonstrated a new technique to obtain cold
atoms in one dimension. \ We have cooled to temperatures as low as
1/35th of the initial temperature of 26 uK, reaching 750 nK \ with
an \ efficiency of 1\%. This technique is not limited by the recoil
temperature and in fact has no fundamental limit. \ This method can
be more efficient than other techniques such as evaporative cooling,
since the efficiency can vary as the square root of the ratio of the
final to initial temperature. \ The cold selected atoms are confined
in a single one-dimensional well, spatially separated from the higher
energy atoms. \ This technique does not produce a thermal distribution
of atoms, but instead imposes a sharp cut in the velocity distribution,
making it useful for studying effects that are sensitive to energy
such as tunneling, chaos and quantum-potential scattering\cite{Raizen,ChaosTunnel,Muga}.

We would like to thank Jung Bog Kim, Hyun Ah Kim and Han Seb Moon
for their help in the laboratory. We thank the Natural Sciences Engineering
and Research Council of Canada, \ the Canadian Foundation for Inovation,
The Ontario Research and Development Challenge Fund and the Canadian
Institute for Photonics Innovation for support of this project.

\bigskip

\bigskip

Fig 1. Principle of velocity selection. a) Atoms in free-space immediately
after release from the trap. b) Potentials turned on. \ Atoms begin
accelerating downhill. \ c) End of selection. \ Cold atoms trapped
in small well while hot atoms are travelling away from the well. \ 

\bigskip

Fig 2. Sample data image of velocity selection to demonstrate spatial
separation of selected atoms from hotter atoms. \ A large potential
gradient was used to show the separation before significant expansion
occured.

\bigskip

Fig 3. \ Plot of experimental data showing temperature versus well
depth. \
The solid circles and hollow diamonds refer to weak ($\beta=0.23)$
and strong ($\beta=1.7)$ potential gradients respectively. \ The
solid and dashed lines are theoretical curves for\ the weak and strong
gradients respectively. The theoretical curves are obtained from simulation
with no free parameters.\ \ Both experimental data sets have slopes
with a temperature of about 1/4 of the height of the barrier. \ The
lowest temperature achieved was 750 nK.

\bigskip\ 

Fig 4. \ Efficiency vs the well depth is shown. \ The solid circles
and hollow squares refer to weak ($\beta=0.23)$ and strong ($\beta=1.7)$
potential gradients respectively. \ The solid and dashed lines are
theoretical curves with no free parameters. 
\end{document}